\documentclass[12pt]{article}

\def\ii{\'{\i}}

\usepackage{epsfig}

\def\menorsim{\smash{\mathop{<}\limits_{\raise3pt\hbox{$\sim$}}}}

\def\maiorsim{\smash{\mathop{>}\limits_{\raise3pt\hbox{$\sim$}}}}

\begin{document}

\title{Percolation and cosmic ray physics above $10^{17}$eV}%
\author{P. Brogueira\footnote{Departamento de F\ii sica, IST, Av. Rovisco Pais, 1049-001 Lisboa, Portugal} ,   
J. Dias de Deus$^{*,}$\footnote{CENTRA, IST, Av. Rovisco Pais, 1049-001 Lisboa, Portugal} ,
 M.C. Espírito Santo$^{*,}$\footnote{LIP, Av. Elias Garcia, 14-1º, 1000-149 Lisboa, Portugal} 
\ and 
M. Pimenta$^{*,\ddag}$
}
\maketitle

\begin{abstract}
We argue that the energy dependence of the average position of the shower maximum, $\bar X_{max}$, and of the number of produced muons, $N_{\mu}$, can be explained by a change, around $10^{17}$eV, in the energy dependence of the inelasticity $K$, which decreases with the energy above the string percolation threshold.
\end{abstract}

\bigskip
\bigskip

The chemical composition of cosmic rays at very high energy, say above $10^{17}$ eV, is a matter of controversy. The reason being that direct measurement is only possible up to $\sim 10^{15}$ eV [1], and a tendency for the effective mass number $A$ to increase with the primary energy $E$ is observed. Data at higher energy, in particular from HiRes experiments [2,3], show that consistency between existing theoretical models and observation requires the effective mass number $A$ to decrease with energy. This may be due to the presence of extra-galactic nucleons [4].

In order to make our argument simple, we shall use the original Heitler idea [5]: the location of the shower maximum is, on the average, related to $\log E, \bar {X}_{max} \sim \log E$, and the number of charged pions or muons is proportional to $E$, $N_{\pi^+ \pi^-} \sim N_{\mu} \sim E^{\beta}, \beta \menorsim 1$. If one has a nucleus with $A$ nucleons colliding in an hadronic collision we shall write,

$$
\bar X_{max} \sim \bar X_0 \log (E/A) ,\eqno(1)
$$
and
$$
\bar N_{\mu} \sim A (E/A)^{\beta} ,\eqno(2)
$$
or
$$
\log N_{\mu} \simeq (1-\beta) \log A +\beta \log E \ . \eqno(3)
$$

We further have, for the elongation length,
$$
{d\bar X_{max} \over d\log E}\simeq \bar X_0 \left[ 1- {d\log A \over d\log E}\right] ,\eqno(4)
$$
and for the $\log N_{\mu}$ dependence on $E$,

$$
{d\log N_{\mu} \over d\log E}=(1-\beta) {d\log A \over d\log E}+\beta \ .\eqno(5)
$$

\noindent As experimentally [2,3], above $10^{17}$ eV $d\bar X_{max} /d\log E$ is larger and $d\log N_{\mu}/$ $d\log E$ is slighty smaller, in comparison with lower energies, the conclusion is:
$$
d\log A/ d\log E < 0 \ , \eqno(6)
$$
i.e., for $E \maiorsim 10^{17}$ eV the average mass number $A$ decreases with energy.

\begin{figure}[t]
\begin{center}
\includegraphics[width=5cm]{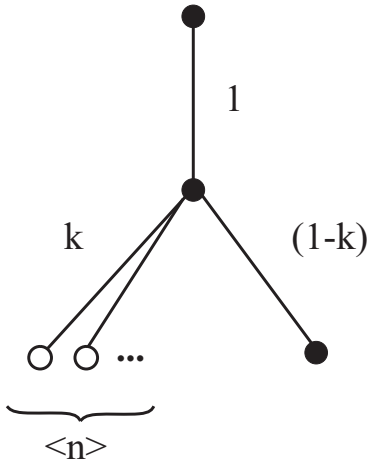}      
\end{center}
\caption{Schematic representation of the first hadronic collision. Most of the energy fraction, $1-K$, is taken by the leading particle. Most of the particles, $\pi^{\pm}$, come from the fraction of energy left, $K$.}
\end{figure}

Another aspect to be taken into account is the $X_{max}$ distribution: it is larger for lighter nuclei (in particular in the case of protons). This is confirmed by Monte Carlo simulations [2], and is seen in data: as the energy increases the $X_{max}$ distribution becomes wider (more proton-like), thus confirming HiRes results. For a detailed and critical discussion see [6].

However, as pointed out in [7], the effect of changing $A$ is equivalent to the effect of changing the average inelasticity $K$. In Fig.1 we schematically present the situation after the first (hadronic) collision. The quantity $(1-K)$ represents the fraction of energy concentrated in the fastest particle, while the inelasticity $K$ is the fraction of energy distributed among produced pions. This means that, in the spirit of Heitler model, we shall have, for $\bar X_{max}$,

$$
\bar X_{max} \simeq \bar X_1 + \bar X_0 \log [(1-K){E\over E_0}] \ ,\eqno(7)
$$
instead of (1), where $\bar X_1$ is the average depth of the first collision, $\bar X_0$ is the radiation length and $E_0$ a low energy threshold. For $\pi^+ ,\pi^-$ or $\mu$ production,

$$
N_{\pi^+ \pi^-} \sim N_{\mu} = C K E \ , \eqno(8)
$$
instead of (2). We further have,
$$
{d \bar X_{max} \over d\log E}\simeq \bar X_0 \left[ {d\log (1-K) \over d\log E} +1\right] \ ,\eqno(9)
$$
instead of (4), and
$$
{d\log N_{\mu} \over d\log E}= {d\log K \over d\log E} + 1 \ ,\eqno(10)
$$

\noindent instead of (5). The condition (6) becomes now 
$$
{d\log K \over d\log E} < 0 \ ,\eqno(11)
$$
i.e., the inelasticity $K$ has to decrease with the energy.

Essentially, all the existing high energy strong interaction models based on QCD, and QCD evolution, predict an increase with energy -- not a decrease -- of the inelasticity $K$ [8]. The same is true for the hadronic generators Sibyll [9] and QGS jet [10], used in cosmic ray cascade analysis. This happens because evolution in the energy implies transfer of energy from valence partons or strings, or from bare Pomeron diagram, to sea partons or strings, or to multi-Pomeron contributions.

\begin{figure}[t]
\begin{center}
\includegraphics[width=12cm]{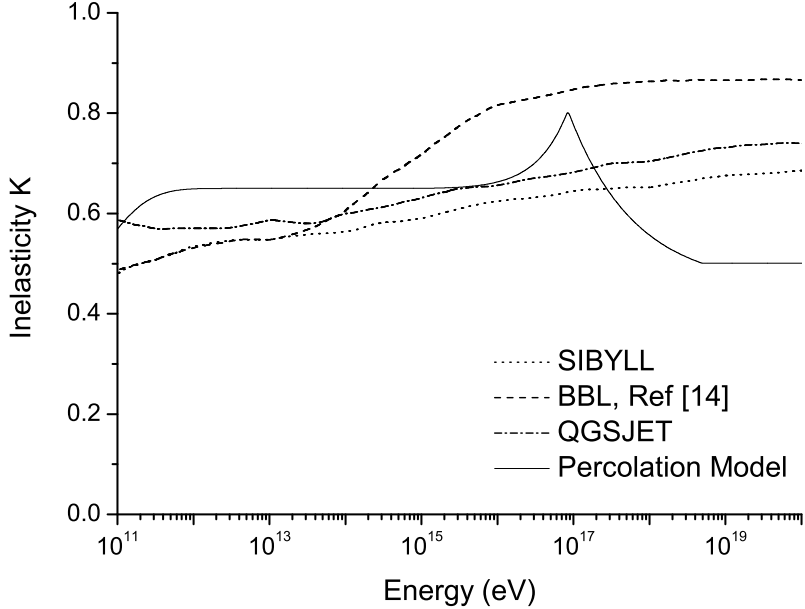}      
\end{center}
\caption{Energy dependence of the average inelasticity $K$. Conventional QCD model results are from [14]: dotted curve is Sibyll simulation; dashed-dotted curve is QGS jet simulations; dashed is a model of [14]; full line is our percolation model.}
\end{figure}

However, in models with percolation of partons or strings, one expects the inelasticity $K$, above the percolation threshold, to decrease with the energy [11]. In the framework of the Dual String Model [13] -- but we believe the argument is more general -- what happens at low energy is the transfer of energy from the valence strings to sea strings (and $K$ increases), while at higher energy the strings start to overlap in the impact parameter and a cumulative effect occurs: the length in rapidity of fused strings is larger [12,11]. At some stage, close to percolation threshold, the percolating strings take over the valence strings, and from then on $K$ decreases with the energy. Percolation is, in fact, a mechanism for generating fast leading particles.

In Fig.2 we show the energy dependence of $K$ in the case of the string percolation model [11], in comparison with $K$ determined from QCD inspired models, without percolation (see, for instance, [14]).

\begin{figure}[t]
\begin{center}
\includegraphics[width=7cm]{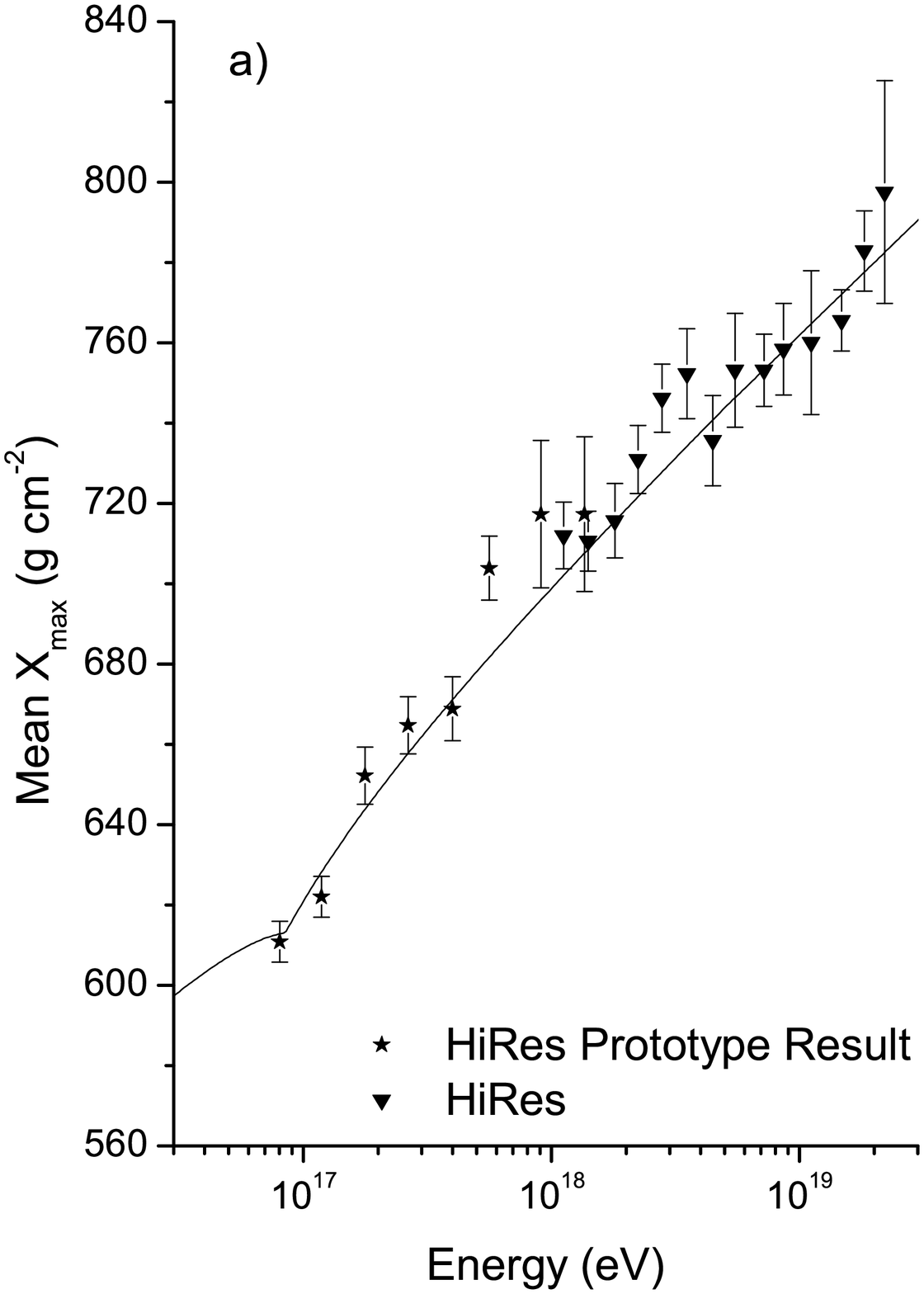}\includegraphics[width=7cm]{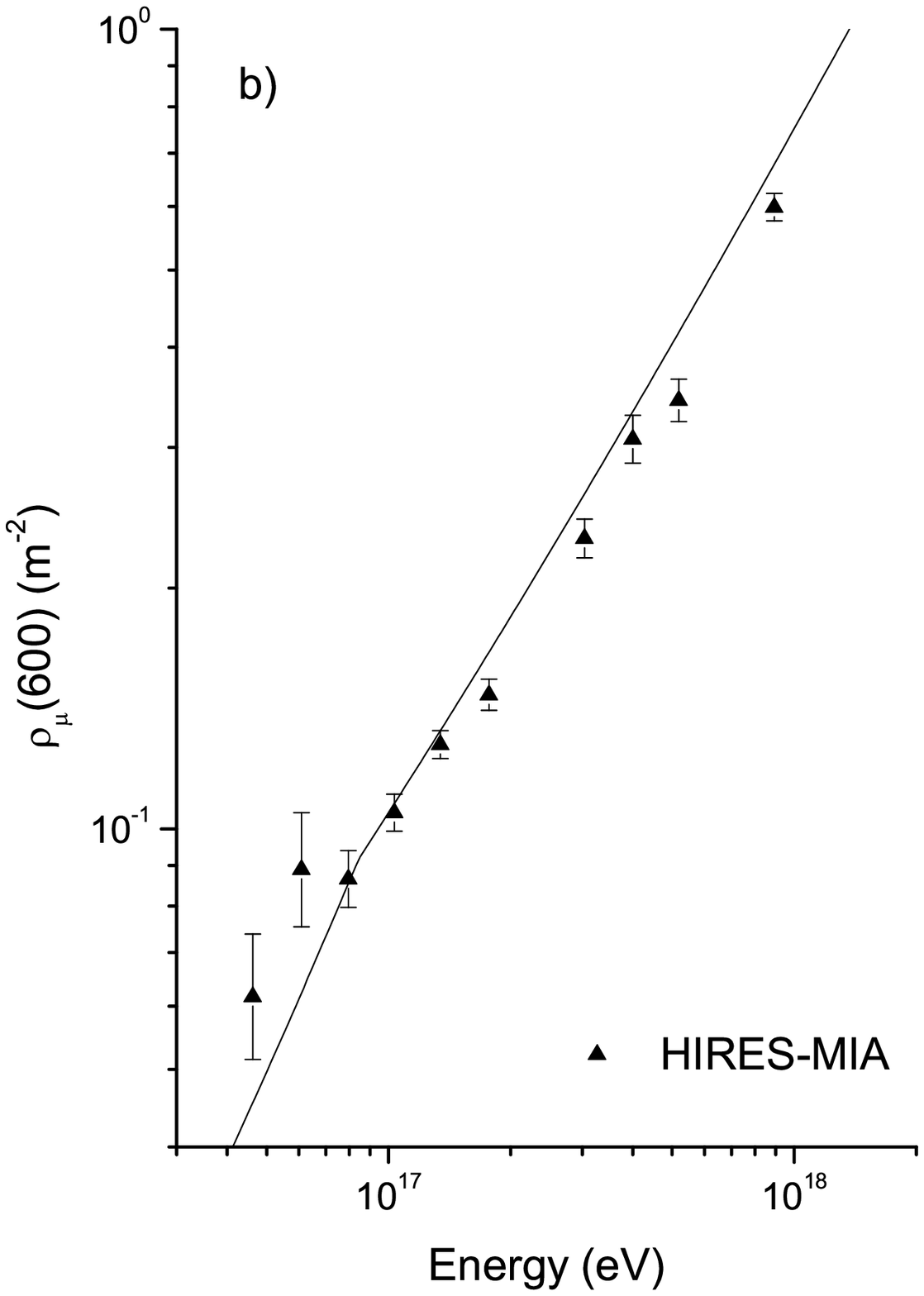}      
\end{center}
\caption{a) Average depth of the shower maximum, $\bar X_{max}$, b) the number density of muons, $\rho_{\mu} (600) m^{-2}$. Data are from [2,3]. Our curves are from (7) and (8) with $\bar X_1 = 60$gcm$^{-2}$, $\bar X_0 =60$gcm$^{-2}$, $E_0=10^7$eV, $C=10^{-17.87}$eV$^{-1}$m$^{-2}$. It is assumed that $\rho_{\mu} (600m)\sim N_{\mu}$.}
\end{figure}

In Fig.3, making use of the values of $K$ for the percolation model of Fig.2, and of Eqs. (7) and (8), we show our results for the average depth of the  shower maximum, $\bar X_{max}$, and for the logarithm of the number $N_{\mu}$ of muons, in comparison with HiRes and HiRes-Mia data. As a consequence of the decrease of $K$, and of Eqs. (9) and (10), there is at $E \maiorsim 10^{17}$eV a fast increase of $\bar X_{max}$ and a slow increase of $\log N_{\mu}$, as seen in data.

It should be mentioned that low inelasticity events generate large values of $X_{max}$ and larger $X_{max}$ fluctuations. This is seen in Fig.4, for $X\equiv X_{max} - X_1$, as a function of event elasticity, taken from [15], by using Sibyll and QGS jet simulations incorporated in the CORSIKA program [16]. One further notes that $X$, at fixed energy, has an approximately $\log$ behaviour, shown in the figure, similar to the behaviour of $\bar X$ as a function of the average inelasticity $K$ (Eq.(7)).

\begin{figure}[t]
\begin{center}
\includegraphics[width=7cm]{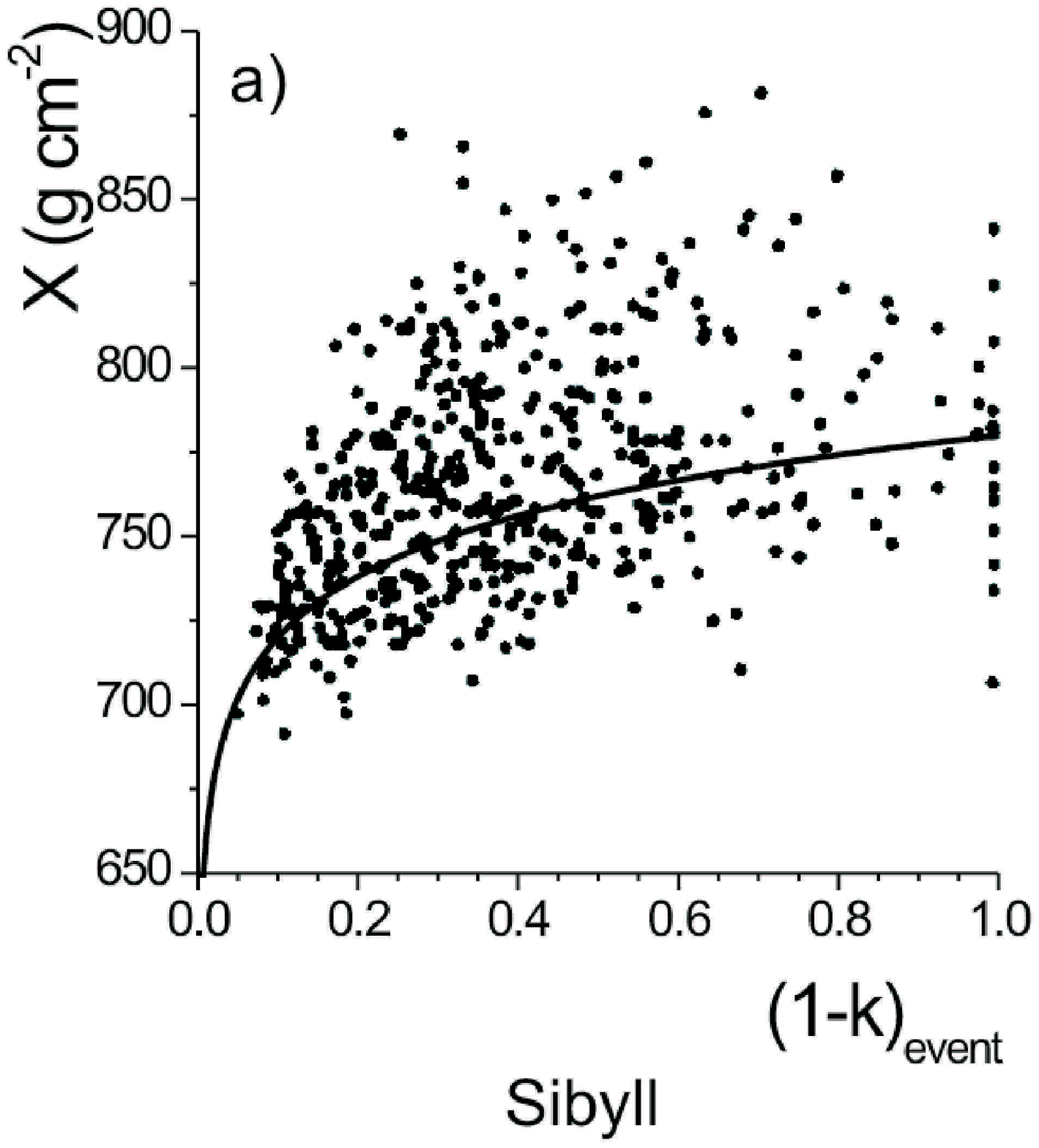}\includegraphics[width=7cm]{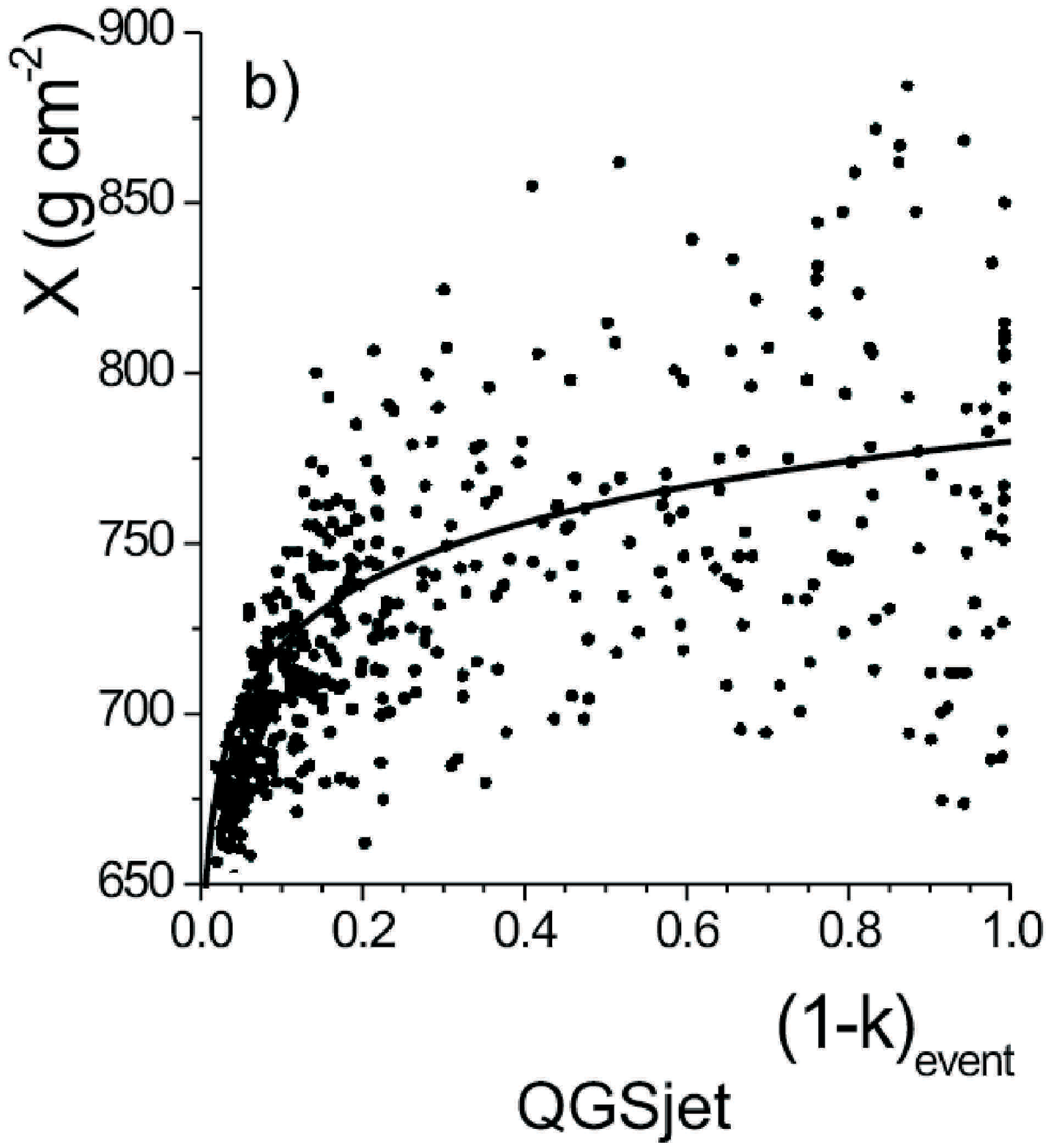}      
\end{center}
\caption{The relative depth $X_{max}-X_1$ as a function of event elasticity $(1-K)$ from [15] by using Sibyll and QGS jet simulations. The curve, from (7), for $E=10^{19}$ eV, shows the positive $X_{max}-X_1$, $(1-K)$ correlation.}
\end{figure}
In conclusion, the effect of a decrease of the inelasticity $K$, for energies of the order of $10^{17}$ eV, is similar to the effect of a decrease of the average mass number $A$, and can explain data: Fig.3. We do not claim that there is not an increase in the fraction of protons in the primary, what we claim is that string percolation, requiring a decrease of $K$, can qualitatively, at least, describe the data around and above $E \sim 10^{17}$ eV.

The reasonable question to ask now is the following: is there a difference between the change in $A$ and the change in $K$? In our simplified approach, mostly based on energy conservation arguments, qualitative differences are not seen. However, in the case of percolation the fast particles are not just protons, they may be fast heavy quark (s,c,b,t) mesons with very different signatures from fast protons. This kind of effect, which amounts to a rapid change of the heavy quark structure functions, can in principle be seen in high energy cosmic rays, as well as in LHC experiments.

We would like to thank Carlos Pajares, Jaime Alvarez-Mu\~niz, Nestor Armesto and J. Guilherme Milhano for discussions. This works has been done under contracts POCTI/36291/FIS/2000 and POCI/FIS/55759/2004 (Portugal).
\bigskip
\bigskip

{\bf References}

\bigskip

\begin{enumerate}
\item T. Shibata, Nucl. Phys. B 75A (1999) 27; K. Asakamori et al., Ap. J. 502 (1998) 278; M.L. Cherry et al, Proc. 26th Int. Cosmic Ray Conference, Salt Lake City, 3 (1999) 163.
\item R. Abbassi et al., Ap. J. 622 (2005) 910.
\item Abu-Zayyad et al., Phys. Rev. Lett. 84 (2000) 4276.
\item J. Linsley and L. Scarci, Phys. Rev. Lett. 9 (1962) 123; Phys. Rev. Lett. 9 (1962) 126.
\item M. Heitler, 1944, Quantum Theory of Radiation, Oxford University Press.
\item A.A. Watson, astro-ph/0410514, 2004, based on a talk at XIII ISVHECRI: Pylos, Greece, 2004.
\item L.W. Jones, Nucl. Phys. B75A (1999) 54.
\item B.Z. Kopeliovich, N.N. Nikolaev and I.K. Potashnikova, Phys. Rev. D39 (1989) 769; A. Capella, U. Sukhatme, C.I. Tan and J. Tran Thanh Van, Phys. Reports 236 (1994) 225; J. Ranft, Phys. Rev. D51 (1995) 64; L. Durand and H. Pi, Phys. Rev. Lett. 58 (1987) 303; H.J. Drescher, A. Dumitru and M. Strikman, hep-ph/0408073; S. Ostapchenko, hep-ph/0501093; I.N. Mishustin and J.L. Kapusta, Phys. Rev. Lett. 88 (2002) 112501. 
\item R.S. Flecher, T.K. Gaisser, P. Lipari and T. Stanev, Phys. Rev. D63 (2001) 054030.
\item N.N. Kalmykov, S.S. Ostapchenko and A.I. Pavlov, Nucl. Phys. Proc. Suppl. 52B (1997) 17.
\item J. Dias de Deus, M.C. Espírito Santo, M. Pimenta and C. Pajares, hep-ph/0507227v2.
\item M.A. Braun, E.G. Ferreiro, F. del Moral and C. Pajares, Eur. Phys. J. C25 (2000) 249.
\item N.S. Amelin, M.A. Braun and C. Pajares, Phys. Lett. B306 (1993) 312; N.S. Amelin, M.A. Braun and C. Pajares, Z. Phys. C63 (1994) 507; M.A. Braun and C. Pajares, Eur. Phys. J. C16 (2000) 349; M.A. Braun and C. Pajares, Phys. Rev. Lett. 85 (2000) 4864, C. Pajares, D. Sousa and R.A. V\'asquez, Phys. Rev. Lett. 86 (2001) 1674.
\item H.J. Dresher, astro-ph/0411143.
\item O.C. Pryke and L. Voyvodic, Nucl. Phys. B (Proc. Suppl.) 75A (1999) 365.
\item J. Knapp et al, Nucl. Phys. B (Proc. Suppl.) 52B (1997) 136; O. Heck et al., Nucl. Phys. B (Proc. Suppl.) 52B 139.
\end{enumerate}
\end{document}